\documentclass[twocolumn, showpacs,
preprintnumbers, amsmath, amssymb, nofootinbib]{revtex4-1}
\usepackage[dvips]{graphicx}

\usepackage{color}

\newcommand{\norm}[1]{\left\lVert#1\right\rVert}
\renewcommand{\vec}[1]{\overrightarrow{#1}}

\begin{document}

\title{
Time-delayed electromagnetic radiation reaction}

\author{Sofiane Faci and Mario Novello}
\affiliation{
CBPF -
Rio de Janeiro - Brazil
} 
\email{sofiane@cbpf.br}

\date{\today}

\begin{abstract}
The Lorentz-Abraham-Dirac (LAD) equation has proved valuable in describing the motion of radiating electric charges but suffers from runaway, pre-acceleration and other ambiguities. The usual scheme is problematic because of locality, which leads to self-interaction with the propagating radiation (i.e. real photons).
Instead, the present heuristic model relies on an infinitesimal time delay between the action of external forces and the inertial reaction by the charge. This yields a new and pathology-free equation of motion whereas the radiated energy-momentum is expressed as an infinite series that generalises Larmor's formula and leads to testable predictions using current and near future ultra-intense lasers.
The time-delay hypothesis is to be put in parallel with recently observed delays of order $10^{-18}$s  (attosecond) in photoemission by atoms and small molecules. Such behaviour is extended here to elementary charges which are supposed to exhibit delays given by the time taken by light to cross the charge's classical radius, which gives $\approx 10^{-23}$s for an electron.
\end{abstract}

\maketitle



\paragraph*{Introduction.}
The advent of ultra-intense lasers provides new possibilities to explore extreme regimes of electrodynamics \cite{di2012extremely, burton2014aspects}.  For example, the forthcoming ELI project is expected to reach intensities up to $10^{24}$Wcm$^{-2}$ \cite{ELI}. At these levels, problems such as radiation reaction \cite{gromes2015lorentz}, nonlinear Compton scattering \cite{hartemann1996classical, koga2005nonlinear} or Breit-Wheeler pair creation \cite{krajewska2012breit} become experimentally  testable. Sophisticated numerical simulations have explored these phenomena based on available theoretical frameworks and have ensued  several interesting results \cite{vranic2016classical, e-trapping, ji2014radiation, green2014transverse}. 
Surprisingly enough, radiation reaction, besides being the oldest, appears to be the most problematic, see \cite{Hammond-2010} for a recent review.  The difficulty lies not in Maxwell's equation but in the equation of motion of a radiating charge. 
The standard equation 
is called Lorentz-Abraham-Dirac (LAD) and reads (we use $c=k=\hbar=1$),
\begin{equation}\label{LAD}
m  \ddot z^\mu = f^\mu + m \epsilon (\dddot {z}^\mu + \ddot z^2\, \dot z^\mu),
\end{equation}
where $\epsilon = \frac{2 \,  e^2}{3m}$, $f$ is an external force and 
$\ddot z^\mu = \frac{d^2}{d\tau^2}z^\mu$ the acceleration; $z^\mu=z^\mu(\tau)$ are the coordinates of the charge worldline given as functions of the proper time $\tau$. 
The LAD equation is known to suffer from pre-acceleration and runaway pathologies. It is also plagued by several ambiguities, ranging from its time-reversibility or not \cite{Rohrlich-1998, Zeh-1999, Rovelli-2004} to the mysterious uniform acceleration case which appears to be in conflict with the Equivalence Principle \cite{Feynman-gravity, harpaz1998radiation}.
In addition it is not clear how to derive the radiated momentum, given by Larmor's formula,
\begin{equation}\label{Larmor}
\delta P^\mu_{Larmor} = - m\, \epsilon \, \ddot z^2\, \dot z^\mu,
\end{equation}
 from the LAD equation \cite{Faci:2016ab, Faci:2016aa}. Indeed,
Larmor's formula (\ref{Larmor}) can be derived by integrating the Li\'enard-Wiechert retarded potential over the whole past history of the charge, which provides a globally defined four-momentum.
However it is not possible to derive it using (\ref{LAD}) (i.e. locally) which yields an energy balance paradox.

\paragraph*{Self-interaction with real photons.}

\begin{figure}[h]
\includegraphics[width = 8.7cm]{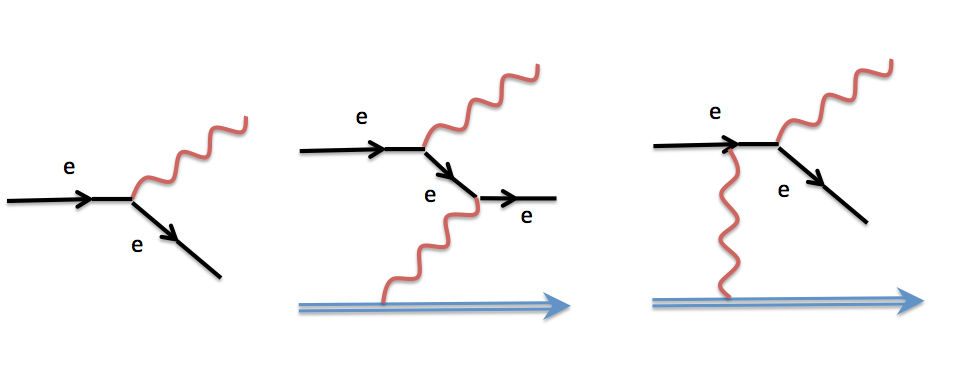}
  \caption{The left diagram represents self-interaction with outgoing radiation, which does not occur in Nature because it violates energy-momentum conservation.
   The middle and right diagrams represent interaction with an external agent through the exchange of (virtual) radiation and leading to emission of (real) radiation and inertial deviation.}
  \label{Fig1}
\end{figure}

We claim the reason behind these difficulties lies in the very hypothesis of (strict) locality which leads to the necessity of self-interaction with the outgoing (real) photon \cite{poisson1999introduction}. A similar problem appears in Gravity \cite{poisson2004motion}.
This is problematic because even in quantum mechanics a charge (massive particle) cannot interact with a photon that it has emitted. What is allowed is self-interaction with virtual photons, never with real photons.
Let us borrow Feynman's diagrams  (in the Furry picture) from QED to represent self-interaction with outgoing radiation. This is depicted in the left part of Fig.~\ref{Fig1} and corresponds to a free electron emitting a photon and being deviated to account for the lost momentum. This process however does not occur in Nature because it violates energy-momentum conservation \cite{jackson1999classical}. Indeed, a free charge cannot emit nor absorb radiation, which is easily inferred by working in the electron rest frame. The physically sensible process is a charge interacting with an external force which is then accompanied by the emission of radiation.

The formulas behind the usual scheme are given by Lorentz's equation of motion
$m\, \ddot z^\mu(\tau) = e\, F^{\mu\nu} \dot z_\nu$ coupled to Maxwell's equation $\partial_\nu F^{\mu\nu} = 4\pi \, J^\mu$. 
The problems appear when $J^\mu$ is given by the charge own current, $J^\mu(z)=e \, \int \dot z^\mu(\tau')\delta(\tau-\tau')d\tau'$. 
That is, the accelerated charge emits radiation that is  in turn causing the acceleration.
On the one hand this naturally leads to divergent field and acceleration, giving the origin of the runaway pathology. 
On the other hand this appears to violate causality, resulting in the pre-acceleration pathology. 
This scheme is misleading because it ignores the external forces that account for half of the full and physically accurate picture. In order to radiate, an unbound charge must be accelerated and for that to happen an external force is necessary, see Fig.~\ref{Fig1}.
Note that Landau and Lifshitz \cite{Landau-Lifshitz} have related the radiation force to $f$ by order reducing the LAD equation, which led them to
$m  \ddot z_\mu = f_\mu +  \epsilon\,  (\dot{f}_\mu + f^2/m\, \dot z_\mu)$. This equation is free from runaway solutions but is limited to slowly varying forces and suffers from the remaining difficulties associated with the LAD equation.
\\

\paragraph*{The time-delay hypothesis.}

We formulate the hypothesis of an infinitesimal time delay between action and reaction. 
This means that the fields interact quasi-locally \cite{tomboulis2015nonlocal} and, in particular, the emitted radiation does not act on its charge; self-interactions are limited to virtual radiation whose effect is to generate a finite mass renormalization. 
The time delay is given by $\epsilon = \frac{2 \,  e^2}{3m}$, which is of order $0.6 \times 10^{-23}$s for the electron and corresponds to $2/3$ the time that takes light to cross its classical radius, $r_c$ (for simplicity, let $r=2r_c/3$, so that $\epsilon=r$). 

The hypothesis can be put in parallel with recently observed time delays in atomic and molecular photoemission \cite{schultze2010delay}. This became possible due to the advent of the so-called attosecond ($10^{-18}$s) chronoscopy which has raised fundamental questions and generated intense theoretical and mostly experimental activity \cite{klunder2011probing, dahlstrom2012introduction, huppert2016attosecond}. In a recent review Pazourek and his colleagues asked: \textit{Is photoionization instantaneous or is there a finite response time of the electronic wave function to the photoabsorption event?}  \cite{pazourek2015attosecond}. The answer is positive and was confirmed by direct observations of attosecond time delays for small atoms and molecules. The theoretical interpretation uses Wigner's work on scattering phase shift and states that electrons take time to \textit{climb} the internal potential before leaving the atom/molecule \cite{wigner1955lower}.

Note that $\epsilon$ is only five orders of magnitude under the present experimental capabilities. It seems rather possible to attain such levels in the near future. A recent proposal has demonstrated the technical possibility of reaching precisions of $10^{-21}$s by using high harmonic x-ray pulses generated with mid-infrared lasers \cite{hernandez2013zeptosecond}.
On the other hand, $r$ (and therefore $\epsilon$) results from direct experiments. Indeed the observed Thomson and also Compton scattering  cross-sections are exactly equal to its square in the forward direction (with a factor $6\pi$).
Accordingly, $r$ (and $\epsilon$) is measurable, Lorentz-invariant and { should thus be seen as a scalar with dimension of length (or time). 
This can also be inferred from the very definition of $r$ (or $\epsilon$) which only depends on the mass and charge of the electron, all intrinsic and Lorentz-invariant properties\footnote{Some confusion might persist with the dubious concept of relativistic vs. rest mass. We refer to Okun \cite{okun2009mass} and Roche \cite{roche2005mass} who have clarified this matter. Energy and momentum are observer-dependent while the mass comes out of the Casimir $P^2=m^2$ and is therefore Lorentz invariant.}.

No further assumptions with respect to the shape or size of the charge particles are needed.
Notwithstanding, one can imagine the electron as a spacelike extension which might be locally represented by the four-vector $\epsilon^\mu=(0, \vec{\epsilon})$, with $\norm{\vec{\epsilon}}=\epsilon$. Thus the invariant scalar reads $\epsilon^\mu \epsilon_\mu = -\epsilon^2$ which is the same for all observers. These see the electron in a different shape but the invariant scale remains the same (the problems related to the rigid spherical electron do not apply to this model). 
Hence the time-delay hypothesis does
 not break Lorentz symmetry\footnote{A breaking usually happens in models requiring some minimal scale. A famous example is Doubly Special Relativity or DSR \cite{amelino2002relativity, magueijo2002lorentz, kowalski2005introduction} which requires fundamental minimal time $t_P$ and length $l_P$. These emerge from several Quantum Gravity models \cite{padmanabhan1985planck, garay1995quantum} but have been afterwards equivocally interpreted as the zeroth and three-vector (norm) components of a four-vector, respectively. Also DSR breaks Lorentz symmetry in its linear representation (the one behind Special Relativity and the Standard Model for instance). This is misleading because it is like working in momentum space and expecting some minimal value for kinetic energy $K=E-m=(\gamma-1)m$ or three-momentum $\vec{P}$ while the Lorentz invariant is the mass $m^2=E^2-\vec{P}^2$.}, or more precisely the isochronous Lorentz group $SO^\dag(1,3)$ which is the homogeneous part of the Poincar\'e group.  Several experiments provide stringent bounds on possible violations \cite{mattingly2005modern, liberati2013tests, kostelecky2016testing, Aaij:2016mos}.
\\

\paragraph*{A new equation of motion.}
In order to implement the time-delay hypothesis, the first attempt is to define the equation, 
$ f_\mu(\tau) = m\,  \ddot z_\mu(\tau+\epsilon)^\perp = m\, \perp_{\mu\nu}\! \ddot z^\nu(\tau+\epsilon),$
 where $\perp_{\mu\nu} = \eta_{\mu\nu} \, - \parallel_{\mu\nu}$ is the projector on the hyperplane $\Sigma(\tau)$ orthogonal to the charge worldline (i.e. to $\dot z^\mu$) at instant $\tau$, with 
$ \parallel_{\mu\nu} =  \dot z_\mu\, \dot z_\nu$ being the parallel projector on the worldline. 
In fact, $\perp$ projects the parallel transported $\ddot z_\mu(\tau+\epsilon)$ on $\Sigma(\tau)$. This is needed for consistency since $ f_\mu(\tau) \in\Sigma(\tau)$ whilst $\ddot z_\mu(\tau+\epsilon) \in\Sigma(\tau+\epsilon)$, the two hyperplanes being not parallel, except for inertial motion, see Fig.~\ref{Fig2}.
However, this equation leads to momentum conservation violation for a negative or vanishing force (i.e. $f_o\leq0$). The underlying reason is its clear asymmetry between force and acceleration, the latter being nothing but a special kind of force, the inertial force \cite{inertia}.
Indeed, for positive motion the momentum flows from the external (potential) sector to the kinetic sector,  that is from $f$ to $\ddot z$. A negative motion should exhibit exactly the opposite flux, momentum flowing from the kinetic sector to the external  sector.  
To settle this asymmetry we amend the above equation through, (let $\delta \ddot z(\tau, \epsilon) = \ddot z(\tau+\epsilon)-\ddot z(\tau)$),
\begin{equation}\label{new-LAD}
f_\mu(\tau) - m\ddot z_\mu(\tau)= m \, \left| \delta \ddot z(\tau, \epsilon)^\perp_\mu \right|,
 \end{equation}
 where $\left| \delta \ddot z(\tau, \epsilon)^\perp_\mu \right| = s\ \delta \ddot z(\tau, \epsilon)^\perp_\mu,$ with $s=\text{sgn}( \delta \ddot z(\tau, \epsilon)^\perp_o)$.
It is important to remark that $\delta \ddot z(\tau, \epsilon)$ can be equivalently replaced by $\delta f(\tau, \epsilon) =  f(\tau)-f(\tau-\epsilon)$  in this equation (and throughout the text) provided the external field is far below the Schwinger critical limit, $E_c = \frac{m^2}{e}$ and the frequency under the limit $\epsilon^{-1}$. Indeed the latter generates electron-positron pair creation whilst $E_c$ sets the limit of linear electrodynamics. $E_c$ is far above current experimental capabilities \cite{Bulanov:2010aa} and therefore demanding $f^2 \ll e^2 \, E_c^2$ is a reasonable restriction.

The new equation of motion (\ref{new-LAD}) is a quasi-local differential 
equation\footnote{This kind of equations are called functional or delay differential equations and have been extensively studied in the literature since many systems in biology or engineering exhibit dependence on past (as well as future in the case of economics for example) history \cite{hale1971functional, burton2014stability}. 
However, to our best knowledge, these techniques have never been applied to fundamental phenomena such as the motion of an elementary charged particle. The work of Bel \cite{llosa1982relativistic} which was later revived by Chicone et al.  \cite{chicone2001delay}, models the radiation reaction through a delay-differential equation
 but the time delay they consider is not related to a single elementary charge but given by the radius of a binary system.}. 
Setting $\epsilon=0$ allows to recover strict locality (i.e. instantaneously reactive electron) and  
equation (\ref{new-LAD}) then reduces to the usual Newton's second law. 
Nonlocal equations are known to lack a proper formulation of the initial value problem and to exhibit acausality due to the presence of past or future delays \cite{tomboulis2015nonlocal}. 
The associated configuration space is infinite-dimensional and contains all sufficiently smooth functions defined in the interval $[\tau_i-\epsilon, \tau_i+\epsilon]$ around some initial time $\tau_i$.
Equivalently, since the time delay $\epsilon$ is constant and infinitesimal (with respect to sensitive time intervals), the above equation can be expanded to an infinite series at a single instant $\tau$, using $\delta \ddot z(\tau, \epsilon) = \sum_{n=1}^{\infty} \frac{\epsilon^n}{n!}\, {z_\mu^{(n+2)}}(\tau)$ with ${z^{(n)}_\mu} = d_{\tau}^n \, z_\mu $ and 
$d_{\tau} = d /d \tau=\dot z^\mu \partial_\mu$,
\begin{equation}\label{new-LAD-expand}
f_\mu(\tau) -  m \ddot z_\mu(\tau) 
= m\, s\, \sum_{n=1}^{\infty} \frac{ \epsilon^n}{n!}\, {z_\mu^{(n+2)\perp}}(\tau),
\end{equation}
hence the initial value problem reduces to set $z_i, \dot z_i$ and all higher order derivatives of the acceleration or the external force. In practice one needs only to know the first few orders to precisely solve the equation.

Up to the first order, equation (\ref{new-LAD-expand}) reduces to 
$$m \ddot z_\mu(\tau) =f_\mu(\tau) - s\, m\epsilon \ \dddot z^\perp_\mu + o(\epsilon^2),$$
 with now $s=\text{sgn}(\dddot z^\perp_o)$.
This is the LAD equation (\ref{LAD}) when $\dddot z_o^\perp<0$, implying $s=-1$, which corresponds for example to circular motion (cyclotron and synchrotron). For $\dddot z_o^\perp>0$ the radiation force has an opposite sign in comparison with the LAD equation and this, in principle, is experimentally testable.
Note that $\dddot z_o^\perp<0$ leads to pre-acceleration behaviour whereas $\dddot z_o^\perp>0$ exhibits post-acceleration. Let us compare with what happens in quantum mechanics, using  Fig.~\ref{Fig1}. The middle diagram stands for typical pre-acceleration since the charge accelerates before interacting with the external force. The diagram on the right side exhibits more intuitive post-acceleration.
Hence, our model fixes the systematic pre-acceleration behaviour of the LAD equation, however, some configurations do inevitably lead to pre-acceleration.
Note finally that the time-irreversal character of our new equation of motion is evident from (\ref{new-LAD}) and further more from (\ref{new-LAD-expand}) for even and odd high-order derivatives do not transform equally under time reversal.

\begin{figure}[h]
\includegraphics[width = 8.5cm]{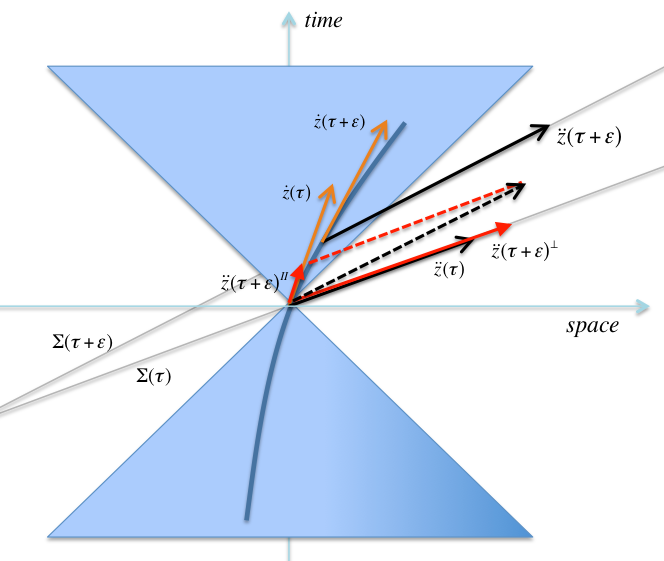}
\caption{
The delayed acceleration $\ddot z^\mu(\tau+\epsilon)$ is parallel transported to the instant $\tau$ and then projected. The orthogonal projection $\ddot z^\mu(\tau+\epsilon)^\perp$ enters the equation of motion (\ref{new-LAD}) while the parallel projection $\ddot z^\mu(\tau+\epsilon)^\parallel$ yields the radiated momentum (\ref{power}).
}
\label{Fig2}
\end{figure}

\paragraph*{The radiated momentum.}

Up to this point only the orthogonal projection of the delayed acceleration, more precisely $\delta \ddot z(\tau,\epsilon)^\perp$, has been used. The parallel projection 
$\delta \ddot z(\tau,\epsilon)^\parallel=\parallel_{\mu\nu}(\tau)\, \delta \ddot z(\tau,\epsilon)^\nu = \delta \ddot z(\tau,\epsilon).\dot z(\tau) \, \dot z_\mu(\tau),$ 
for being orthogonal to $\Sigma(\tau)$, cannot enter the equation of motion (Fig.~\ref{Fig2}). 
We state that $\delta \ddot z(\tau,\epsilon)^\parallel$ is $cut$ from the charge particle and radiated in electromagnetic form. 
The radiated momentum is thus defined through
\begin{equation}\label{power}
\delta P_{rad}^\mu 
= m \, \delta \ddot z(\tau, \epsilon)^\parallel
= m\, \sum_{n=1}^{\infty} \frac{\epsilon^n}{n!}\, {z_\mu^{(n+2)\parallel}}(\tau).
\end{equation}
Like the equation of motion (\ref{LAD}), this formula is time-irreversible, which is easy to check.
 %
Furthermore it can be put in the form $\delta P_{rad}^\mu= \delta P_{rad} \ \dot z^\mu(\tau)$, where
(using the constraint $\dot z^2=1$), 
$$
\frac{\delta P_{rad}}{m} = - \epsilon\,  \ddot z^2 + \frac{3\epsilon^2}{2} \ddot z.\dddot z  - \epsilon^3\, ( \frac{\dddot z^2}{2} 
+ \frac{2\ddot{\ddot{z}}.\ddot z}{3}) + o(\epsilon^4).
$$
The first term corresponds to Larmor's formula (\ref{Larmor}). 
Higher-order terms are new and do change the behaviour of electric charges in the presence of intense or rapidly varying external forces, as in high-frequency lasers. This provides the most easily testable prediction of our model.

Now, the acceleration vector being spacelike, the Larmor term (zeroth component) is positive. However higher orders can apparently be positive or negative. 

\emph{Proposition: All terms in (\ref{power}) with odd higher derivative are positive (i.e. have positive zeroth components).}

To prove this proposition we work in an arbitrary inertial frame, where one has $\dot z^\mu=\gamma (c,\vec{v})$,  before taking the limit to the instantaneously comoving inertial frame, $\dot z^\mu\to (1, \vec{0})$. Therefore $ { z^{(n)}_\mu} = ({ \gamma^{(n-1)}},  {\gamma^{(n-1)} \vec{v}}) |_{\vec{v}=0}$ and $ z^{(n)}.\dot z =  \gamma^{(n-1)} |_{\vec{v}=0} $. The Lorentz boost function $\gamma = \frac{dt}{d\tau} = 1/\sqrt{1-\vec{v}^2}$ is analytic, positive and even (time-reversible). This means that there exists an infinite set of positive constant coefficients $\{c_i\}$ such that $\gamma(\tau) = \sum_{i=0}^\infty c_{2i}\, \tau^{2i}$.
Thus the odd higher derivatives are also positive. In particular, this holds when $\vec{v}=0$ and therefore $z^{(2n+1)}.\dot z =  \gamma^{(2n)}{ } |_{\vec{v}=0} $ is positive for all $n$. In addition since $ z^{(2n+1)}.\dot z =  \gamma^{(2n)}{ } |_{\vec{v}=0} $ is  scalar the positivity holds for all frames. 

This means that all odd derivatives of the acceleration lie in the half of the space-time delimited by the $\Sigma(\tau)$ hyperplane and containing the velocity vector $\dot z^\mu$.  Hence odd terms are always removing energy from the system (force + charge) and their effect on the motion is naturally resistive or dissipative, which is consistent with them -- as well as equations (\ref{LAD}) and (\ref{power}) -- being time-irrevesible.
The even derivative terms in (\ref{power}) have an indefinite sign and are time-reversible. 
However, within the validity limit of our model, that is $E_c$ for the fields and $\epsilon^{-1}$ for the frequencies, the dominant term in the series (\ref{power}) is the Larmor term and so the radiated momentum is always positive and forward oriented. In addition, performing a motion back and forth results in a null effect on the total amount of radiated energy coming from even terms.
 %
\\

\paragraph*{Energy-momentum conservation.}
Completing the above derivation of the radiated momentum, we show in this part how the present formalism allows to cure the energy balance paradox present in the LAD model.
Using the identity
$$ \delta \ddot z(\tau,\epsilon)^2= (\delta \ddot z(\tau,\epsilon)^\perp)^2 + (\delta \ddot z(\tau,\epsilon)^\parallel)^2,$$
together with equations (\ref{new-LAD}) and (\ref{power}), defining the total momentum flux (between the instants $\tau$ and $\tau+\epsilon$) as 
$\delta P_{tot}^\mu(\tau) = m \, \delta \ddot z(\tau,\epsilon)$ 
and the internal momentum flux as $\delta P_{int}^\mu(\tau)= f^\mu(\tau)- m \ddot z^\mu(\tau)$, 
one reaches the relation
\begin{equation}\label{conservation}
 \delta P_{tot}^2 = \delta P_{int}^2 + \delta P_{rad}^2.
\end{equation}
This formula encodes the conservation of energy-momentum.
It says that the total momentum,  $\delta P_{tot}$ is split into an internal flux $\delta P_{int}$, which flows between the kinetic and potential sectors, and an external flux $\delta P_{rad}$, which is being dissipated.
Moreover since it involves scalar quantities, the relation (\ref{conservation}) holds true in all frames.
\\

\paragraph*{Uniform proper acceleration.}
Let us now consider a practical example, a charge undergoing uniform acceleration, an important case for which the LAD model difficulties become evident.
First of all, note that uniform acceleration  is a  frame-dependent notion for it is not possible to impose $\dddot z^2=0$ when $\ddot z^2\neq 0$, which can be inferred from the identity $\dddot z.\dot z=-\ddot z^2$.
The common treatment in the literature -- usually without recognising or stating it -- deals with uniform acceleration in the rest (instantaneously comoving inertial) frame. Even though this is physically distinguished it remains a particular frame.
Yet this can be put in a covariant form as $\dddot z^\mu = -\ddot z^2 \dot z^\mu$ and $\ddot z^2=constant$. Higher orders read $\ddot z_\mu^{(2n)} = \phi^{2n}\, \ddot z_\mu$ and $\ddot z_\mu^{(2n+1)} = \phi^{2n+1}\, \dot z_\mu$, where $\phi = \sqrt{- \ddot z^2}>0$.
Accordingly, the equation of motion (\ref{new-LAD}) reduces to
\begin{equation}\label{uniform}
 f_\mu -  m\, \ddot z_\mu
 = s \left\{ {\epsilon^2\phi^2}/{2!}  +  {\epsilon^4\phi^4}/{4!} +  \dots \right\} m {\ddot{z}}_\mu,
\end{equation}
where $s=\text{sgn}(\ddot z_o).$ 
The case of positive acceleration, $\ddot z_0>0,$ leads to $f^\mu =  \cosh{(\epsilon \phi)} \,m \ddot z^\mu$ and thus $| f| > | \ddot z |$,
whilst for negative acceleration, $\ddot z_0<0$, one finds $f^\mu = (2-\cosh{(\epsilon \phi)}) \,m \ddot z^\mu$ and thus $| f| < | \ddot z |$. In other words, both the acceleration and the external force are constant but they are not equal, the difference being the source of the radiated momentum. 
The latter reads from (\ref{power}) and yields $\delta P_{rad}^\mu = \delta P_{rad} \, \dot z^\mu$, with $\delta P_{rad} = m\phi \sinh{(\epsilon \phi)}= \delta P_{Larmor}\times \frac{\sinh{\epsilon \phi}}{{\epsilon \phi}}$. The radiated power ($\mu=0$) is larger than predicted by Larmor's formula which is reached in the limit of small acceleration.
As to the radiated three-momentum, $\delta \vec{P}_{rad} =0$, which is interpreted by stating that the radiation is isotropic in the charge frame.
Finally, energy-momentum conservation can be straightforwardly inferred from (\ref{conservation}), (\ref{uniform}) and $\delta P_{rad}$. The square of the total momentum flux then reads
$
\delta P_{tot}^2 = - 2 m^2 \, \phi^2 (1-\cosh(\epsilon \phi)).
$
\\

\paragraph*{Uniform acceleration in the lab frame.}

Another possibility is to define uniform acceleration with respect to the lab frame. 
Let us first write the equation of motion (\ref{new-LAD}) in an inertial frame, $\dot z^\mu=\gamma (c,\vec{v})$,
\begin{equation}\label{LAD-3d}
  {a}.{v} \, \vec{v} + \frac{\vec{a}}{\gamma^{2}}  =  \frac{\vec{f} }{m\, \gamma^3}  - \epsilon \, s \, \gamma^3 \, ( {h}.{v} \, \vec{v} +\frac{\vec{h}}{\gamma^{2}} ) + o(\epsilon^2),
\end{equation}
where $\vec{h} = 3 a.v\, \vec{a} + \frac{\vec{\dot a}}{\gamma^2}$ and $s=\text{sgn}(h.v)$. The zeroth component of (\ref{new-LAD}) is obtained by projecting (\ref{LAD-3d}) on $\vec{v}$.
Uniform acceleration with respect to the lab frame reads $\dot{\vec{a}}=0$. In this case $s=\text{sgn}(h.v) = \text{sgn}((a.v)^2)=+1$. The equation of motion (\ref{LAD-3d}) reduces to 
$m\, \gamma^3\,  \vec{a} =  \vec{f}  -  3 \epsilon \, m\, \gamma^6 \, a^2\,  \vec{v}+ o(\epsilon^2).$ 
In order to maintain $\vec{a}$ uniform (in the lab frame) the external force must vary as $\gamma^3$. Consequently this is very different from the previous uniform proper acceleration case.
As to the power radiated, it reads  $\delta P_{rad}=m\, \epsilon \gamma^6 {a}^2 + o(\epsilon^2)$. 
It is clear that this could not be extracted from the equation of motion. Note finally that in the non-relativistic limit the radiation reaction vanishes whilst the radiated momentum does not, which leads back to the uniform proper acceleration case.
%

\paragraph*{Final remarks.}
We have introduced the hypothesis of a time delay $\epsilon$ between action and reaction to treat the motion of an unbound charge. After defining, $\delta \ddot z(\tau, \epsilon)$, the total momentum flux in the interval $[\tau, \tau+\epsilon]$, the equation of motion (the radiated momentum) reads from its orthogonal (parallel) projection on the charge worldline.
This results in (\ref{new-LAD}), which extends and fixes the LAD equation of motion, and in a new formula for the radiated momentum (\ref{power}). 
Pre-acceleration is unavoidable but the systematic pre-acceleration behaviour of the LAD equation is  fixed here for post-acceleration is possible as well.
Both equations are time-irreversible and hence, if the present model is confirmed,  classical Electrodynamics turns out to be fundamentally irreversible.
This model can be used to build a quasi-local QED with $\epsilon$ setting a natural UV cutoff. Indeed, Feynman's regularization scheme relies on replacing the delta distribution by a Gaussian of width $\epsilon$ \cite{feynman1948relativistic}.
The novelty here is that there is no need to get rid of $\epsilon$ since it acquires a clear physical significance and must be recovered in the classical limit. In other words there is no need for renormalization since no divergencies need to be cured.

For future developments, we are investigating how to rewrite this work in the Lagrangian formalism so that exploring quantum effects will be easier. 
Furthermore, it might be interesting to apply the present model to binary electromagnetic as well as gravitational systems to study higher multipole radiation. 
It might also be applied to other dissipative phenomenas such as the elusive friction.  In particular, friction at the level of a single atom sliding over some hard surface has been observed since the 1990's and led to nano-friction science \cite{ternes2008force}.
More precise measurements have been performed recently and yet a well-matching theoretical model is still lacking \cite{kim2009nano, mo2009friction, gangloff2015velocity}. Furthermore, due to recent observations of time delays in atomic interactions \cite{klunder2011probing, dahlstrom2012introduction, huppert2016attosecond,pazourek2015attosecond}, equations similar to (\ref{LAD}) and (\ref{power}) might lead to new insights into nano-friction.
%
\\

\paragraph*{Acknowledgment.}
\textit{We would like to thank Jos\'e A. Helayel-Neto, Nelson Pinto-Neto, Erico Goulart, Felipe T. Falciano, Marc Casals, Arthur Scardua, Samuel Colin, Miguel Quartin and Sergio Joras for helpful discussions and Satheeshkumar VH, Junior Toniato and Irina Nasteva for reading and improving the manuscript. We also thank CNPq for financial support.}

\baselineskip=10pt
\bibliography{Biblio}

\end{document}